\documentclass[%
 reprint,
superscriptaddress,
 amsmath,amssymb,
 aps,
 prl,
floatfix,
a4paper,
]{revtex4-1}

\usepackage{graphicx}
\usepackage{dcolumn}
\usepackage{bm}
\usepackage{xspace}
\usepackage{color}
\usepackage{ulem}
\usepackage{tabularx}

\usepackage{hyperref}

\graphicspath{{./images/}}

\newcommand{\LCDM}{$\Lambda$CDM\xspace}

\newcommand{\virg}[1]{`{#1}'}

\newcommand{\Mh}{M_{\rm h}}

\newcommand{\Rhl}{R_{1/2}}

\newcommand{\hMsol}{h^{-1}\,{\rm M_\odot}}

\newcommand{\mss}{{\rm m \, s^{-2}}}

\newcommand{\gbar}{g_{\rm bar}}
\newcommand{\gtot}{g_{\rm tot}}
\newcommand{\gcrit}{g_{\dag}}
\newcommand{\sint}{\sigma_{\rm int}}

\newcommand{\aap}{Astron Astrophys}
\newcommand{\apjl}{Astrophys J Lett}
\newcommand{\apjs}{Astrophys J Suppl Ser}

\newcommand{\mnras}{Mon Not R Astron Soc}

\begin{document}

\preprint{APS/123-QED}

\title{On the radial acceleration relation of $\Lambda$CDM satellite galaxies}

\author{Enrico \surname{Garaldi}}
 \altaffiliation{Member of the International Max Planck Research School (IMPRS) for Astronomy and Astrophysics at the Universities of Bonn and Cologne}
\affiliation{Argelander Institut f\"ur Astronomie, Auf dem H\"ugel 71, Bonn, D-53121, Germany.}
\email{Electronic address: egaraldi@uni-bonn.de}
\author{Emilio \surname{Romano-D\'{\i}az}}
\affiliation{Argelander Institut f\"ur Astronomie, Auf dem H\"ugel 71, Bonn, D-53121, Germany.}
\author{Cristiano \surname{Porciani}}
\affiliation{Argelander Institut f\"ur Astronomie, Auf dem H\"ugel 71, Bonn, D-53121, Germany.}
\author{Marcel S. \surname{Pawlowski}}
\altaffiliation{Hubble Fellow}
\affiliation{Department of Physics and Astronomy, University of California, Irvine, CA 92697, USA.}

\date{\today}

\begin{abstract}
The radial acceleration measured in bright galaxies tightly correlates with that
generated by the observed distribution of baryons, a phenomenon known as the radial acceleration relation (RAR). Dwarf spheroidal satellite galaxies have been recently found to depart from the extrapolation of the RAR measured for more massive objects but with a substantially larger scatter.
If confirmed by new data, this result provides a powerful test of the theory of gravity at low accelerations that requires robust theoretical predictions.
By using high-resolution hydrodynamical simulations, we show that, 
within the standard model of cosmology ($\Lambda$CDM), 
satellite galaxies are expected to follow the same RAR as brighter systems but with a much larger scatter which
does not correlate with the physical properties of
the galaxies.
In the simulations, the RAR evolves mildly with redshift. 
Moreover,
the acceleration due to the gravitational field of the host has no effect on the RAR. This is in contrast with the External Field Effect in Modified Newtonian Dynamics (MOND) which causes galaxies in strong external fields to 
deviate from the RAR. This difference between $\Lambda$CDM and MOND offers a possible way to discriminate between them.
\end{abstract}

\maketitle

{\it Introduction.{\bf ---}}The standard $\Lambda$CDM model of cosmology relies on the theory of general relativity and assumes that the energy budget of the universe
is dominated by cold dark matter and a cosmological constant.
The cosmic microwave background, gravitational lensing, and galactic dynamics provide abundant evidence for mass discrepancies which are usually interpreted 
as manifestations of particle dark matter (DM).
However,
its basic constituents have so far eluded direct detection.
Furthermore, tight empirical relations are observed between the luminous and dark
components of galaxies 
\cite{Faber+Jackson1976, Tully+Fisher1977,McGaugh2004,Sancisi2004}.
These remarkable and intriguing correlations might appear `unnatural' in the $\Lambda$CDM model. For this reason, some authors elevated them to fundamental
laws of Nature and developed alternative scenarios without DM.
In the theory of Modified Newtonian Dynamics (MOND) \cite{Milgrom1983}, for instance, 
the observed acceleration $a$ is
given by $a \, \mu(a/a_0)=a_{\rm N}$, where $a_{\rm N}$ is the Newtonian
acceleration, $a_0$ is a new fundamental constant of Nature, and $\mu$ is
an interpolation function such that $\mu\to 1$ for $x\gg 1$ and
$\mu \to x$ when $x\ll 1$. In the
non-relativistic case, the MOND equation can be achieved by changing
either the Newton's second law (modified inertia, \cite{Milgrom1994}) or the Poisson's
equation (modified gravity, \cite{Bekenstein+1984}).

The debate was recently revived when
\cite{McGaugh+2016} and \cite{Lelli+2017}
concluded 
that the (centripetal) radial acceleration ($\gbar$)
generated by the visible baryonic matter in galaxies
and the actual (centripetal) radial acceleration derived from kinematic measurements ($\gtot$) strongly correlate over the range $10^{-12}<\gbar<10^{-8} \, \mss$.
In terms of the characteristic acceleration $\gcrit=
[1.20 \pm 0.02 \,\mathrm{(rnd)}\,\pm 0.24\, \mathrm{(sys)}] \times 10^{-10}$ m s$^{-2}$, the 
spatially-resolved data for 240 galaxies of different sizes and morphological types
scatter around the mean radial acceleration relation (RAR)
\begin{equation}
\label{eq:rar}
\gtot = \frac{\gbar}{1-e^{-\sqrt{\gbar/\gcrit}}} \;,
\end{equation}
i.e. $\gtot \simeq \gbar$ for $\gbar\gg \gcrit$
while $\gtot \simeq \sqrt{\gbar\gcrit}\gg \gbar$ 
for $\gbar\ll \gcrit$.
Eq.~(\ref{eq:rar}) is inspired by
the interpolation function of MOND 
and the existence of the RAR
could be invoked as direct evidence for this alternative theory of gravity 
(basically, the empirical parameter $\gcrit$ embodies $a_0$).
However,
numerical simulations of galaxy formation in the \LCDM framework
reproduce the overall shape of the observed correlation \cite{Santos-Santos+2016,DiCintio+2016,Keller+Wadsley2017,Ludlow+2017} (see, however, 
\cite{Tenneti+2018} for an exception).
Here, the
RAR emerges from the dissipative collapse of baryons within DM halos and is less influenced by
the feedback 
of stars and active galactic nuclei. For disc galaxies forming at the centre of their host halos (central galaxies), the RAR reflects: i) the narrow range of the host virial masses; ii) the self-similar acceleration profiles of CDM haloes; iii) the tight correlation between baryonic mass, galaxy size and halo mass \cite{DiCintio+2016, Navarro+2017}.
However, simulated RARs tend to overpredict the value of $\gcrit$ 
regardless of the adopted subgrid feedback model (except possibly \cite{Keller+Wadsley2017}).
Furthermore, the 
scatter around the RAR for late-type galaxies ($\lesssim 0.13$ dex) is dominated by observational
uncertainties, which is difficult to reconcile with simulations which show an intrinsic spread of comparable magnitude \cite{Desmond2017}.

This Letter focuses on the low $\gbar$ regime which has the
potential to distinguish between the two competing scenarios described above. 
By analyzing a set of satellites of Andromeda and the Milky Way,
\cite{Lelli+2017} found that 
dwarf spheroidal galaxies (dSphs) do not follow Eq.~(\ref{eq:rar}) 
if $\gcrit$ is chosen to fit the data for more massive objects.
Instead of dropping as $\gtot \propto \sqrt{\gbar}$, the total acceleration
stays approximately constant, $\gtot\simeq 10^{-11}$ m s$^{-2}$,
for $\gbar \lesssim 9\times 10^{-12}$ m s$^{-2}$.
It is currently impossible to draw conclusions based on this 
finding. In fact, the expected signal in $\Lambda$CDM has only been computed
for central galaxies that probe larger accelerations than faint dSphs. 
Moreover, as extensively discussed in \cite{Lelli+2017}, it is still unclear whether the observed flattening of the RAR is physical or due to observational artifacts. The inferred masses (or, equivalently,
the values of $\gtot$) for faint dSphs are based on velocity-dispersion measurements \citep{Wolf+2010} and are plagued by considerably larger uncertainties than measurements of rotation curves for late-type galaxies. 
Since dSphs have low velocity dispersions and their estimates are often based on a handful of observable stars, current results might be severely affected by unresolved binary systems \cite{McConnachie+2010}.
Both this effect and out-of-equilibrium dynamics tend to 
inflate the measured velocity dispersions \cite{McGaugh+2010}.

This situation provides us with a unique opportunity to predict the expected behavior of the RAR for satellite galaxies in the $\Lambda$CDM scenario. 

{\it Numerical simulations.{\bf ---}}We use the ZOMG hydrodynamical simulations that have been comprehensively described in 
\cite{zomgI,zomgII,zomgIII}.
These runs follow the process of galaxy formation zooming in on a set of DM haloes with masses $\Mh \approx 3\times10^{11}\,\hMsol$, where $h$ denotes
the present-day value of the Hubble parameter in units of 100 km s$^{-1}$ Mpc$^{-1}$. 
The background cosmology and the linear power spectrum of density perturbations match the best-fit {\it Planck}+WP+highL+BAO model in \cite{Planck2013_Cosmology}.
The mass resolution is 
$m_*=m_{\rm gas}/2 = m_{\rm DM}/10.8 = 1.21 \times 10^4 \, \hMsol$ for stars, gas and DM, respectively. The simulations employ a supernova-feedback model
and the resulting central galaxies closely match the stellar mass-halo
mass and stellar mass-star formation rate relations observed at redshift $z=0$ \citep{zomgII}.
Similarly, the satellite galaxies are consistent with the observed baryonic Tully-Fisher relation, subhalo mass function, stellar fraction and stellar velocity dispersion \citep{zomgIII}.

\begin{figure}
	\includegraphics[width=\columnwidth]{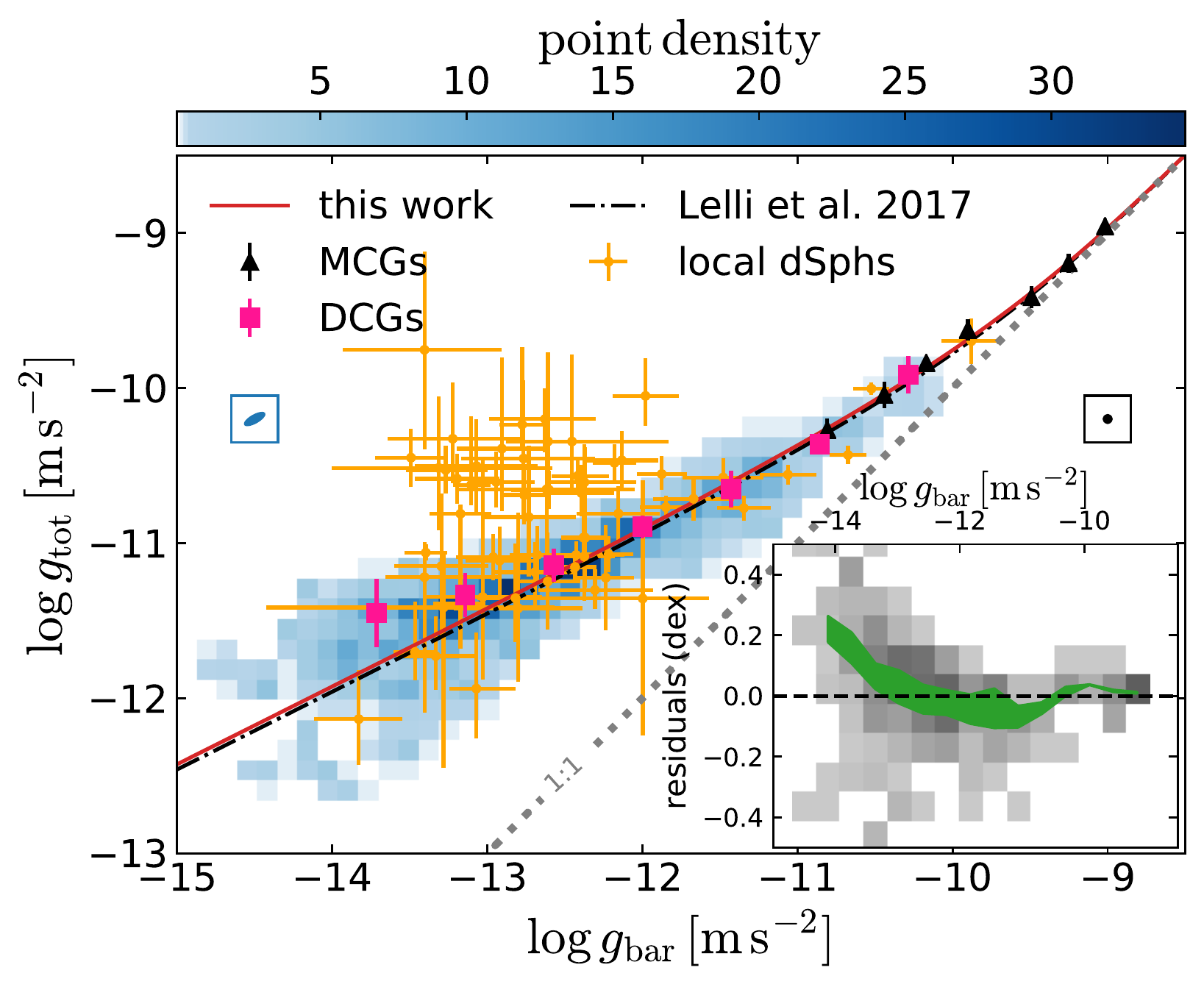}
    \caption{Distribution of observed and simulated galaxies in the $\log\gbar$-$\log\gtot$ plane. 
Triangles and squares indicate the median $\gtot$ in bins of $\gbar$ for the simulated
MCGs and DCGs, respectively (errorbars enclose the central 68\% of the data).
The solid and dot-dashed curves are the best-fit RARs inferred from the MCGs and the observations.  
The large crosses represent the measurements for local dSph satellites presented in 
\cite{Lelli+2017}.
The colored map displays the
number density of the simulated MSGs.
Each object corresponds to a bivariate
Gaussian distribution reflecting the statistical errors.
The framed ellipses show the typical 68\% bootstrap region for objects with $\gbar<10^{-13}\,\mss$ (left) and $\gbar>10^{-10.5}\,\mss$ (right). 
The inset shows the density of the residuals between the MSGs and the best-fit RAR for the MCGs.
The solid band is centred on the mean residual at fixed $\gbar$ and has width equal to the mean measurement error for $\gtot$.}
    \label{fig:rar_z0}
\end{figure}

{\it Method.{\bf ---}}DM haloes and their substructures are identified using the {\sc amiga halo finder} code \cite{AHF1,AHF2}. We associate a `main central galaxy' (MCG)
with each of the resimulated central DM halos by simply considering a spherical region extending for 10\% of the halo radius. 
All substructures with a stellar component
that lie within the splashback radius of the main halo (identified with the abrupt steepening of the spherically averaged mass-density profile as in \cite{Diemer+2014}) are labelled as `main satellite galaxies' (MSGs). 
Finally, we consider the dwarf central galaxies (DCGs) associated with less massive DM clumps
lying between one and three splashback radii from the main halos.
The centripetal accelerations are evaluated as $g_{\rm x} = G \, M_{\rm x}(<r)/r^2$, where $G$ is the gravitational constant and $M_{\rm x}(<r)$ denotes the galaxy mass (total or baryonic) contained within the radius $r$.
For MCGs, 
we compute the acceleration radial profiles and their correlated bootstrap errors (consistent with Poisson fluctuations) at 7 different positions 
extending from 1\% to 10\% of the halo radius equally spaced in log scale.
We find that the resulting $\gtot$ is consistent with measurements based on the gas rotational velocity, as done in observational studies.
For MSGs and DCGs,
accelerations are only computed
at the stellar half-mass radius $\Rhl$ (i.e. the radius within which half of the stellar particles are located) to mimic the half-light radius used for observational data.
We only consider galaxies containing more than 10 (gravitationally bound) stellar particles within
$\Rhl$. 
The covariance matrix for
$\gbar$ and $\gtot$
is
estimated with the bootstrap method by resampling stellar particles within the
individual objects. 
We find that errors on 
$\log \gbar$ and $\log \gtot$ approximately follow a bivariate Gaussian distribution. 
We fit Eq.~(\ref{eq:rar}) to our simulated data.
Using Bayesian statistics, we jointly constrain $\gcrit$ and $\sint$,
the intrinsic scatter around the RAR at fixed $\gbar$ (i.e. the rms value of the residuals of $\log \gtot$).
For each measured pair $(\log\gbar,\log\gtot)$, we consider a Gaussian (partial) likelihood function and we marginalize it over the unknown true value of the bayonic acceleration (which does not coincide with $\gbar$ due to measurement errors).
We write the variance of $\log \gtot$ at fixed $\gbar$ as the sum in quadrature of the measurement error and $\sint$.
Eventually, we build posterior distributions for the model parameters by uniformly sampling the parameter space and assuming flat priors on $\gcrit$ and $\sint$.

\begin{figure}
	\includegraphics[width=\columnwidth]{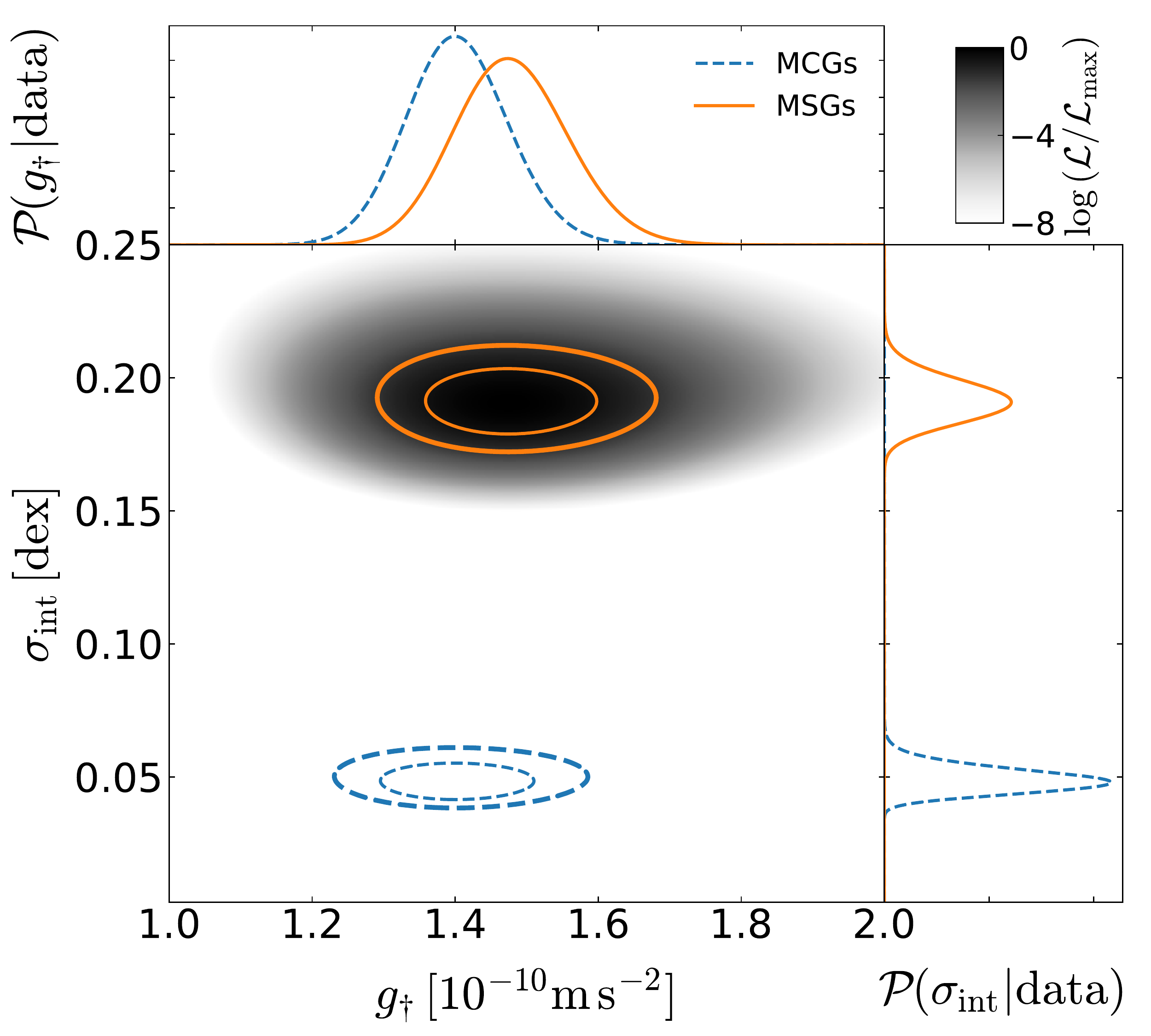}
    \caption{The central image shows the likelihood of the fitting parameters $\gcrit$ and $\sint$ given the simulated MSGs. 
    The solid curves indicate the contour levels enclosing 68\% and 95\% of the posterior probability.
    Fitting the simulated MCGs, instead, produces the dashed contours.
    The top and right panels show the marginalized posterior distributions for $\gcrit$ and $\sint$, respectively.}
    \label{fig:likelihood}
\end{figure}

{\it The RAR at redshift zero.{\bf ---}}Fig.~\ref{fig:rar_z0} compares
real and simulated
galaxies in the $\log\gbar$-$\log\gtot$ plane at $z=0$. 
Our MCGs and DCGs follow a tight RAR which is in excellent agreement with observations. 
For $\gbar < 10^{-12}\,\mss$,  DCGs depart from Eq.~(\ref{eq:rar}) and tend to have higher $\gtot$ (see also \cite{Keller+Wadsley2017,Ludlow+2017}).
The dSph satellite galaxies analyzed in Ref. \cite{Lelli+2017} sprinkle around 
$\gtot\simeq 10^{-11}\,\mss$ independently of $\gbar$.
Conversely, the simulated MSGs form a well defined sequence to a great extent aligned with the observed RAR (but with a larger scatter) and do not show any transition to a constant $\gtot$ for the least massive satellites. 
For $\gbar<10^{-13}\,\mss$, 
their mean $\gtot$ at fixed $\gbar$ lies slightly above the observed RAR of the central galaxies (in fact $\gtot \propto \gbar^{0.4}$ in this regime) but slightly below that of DCGs.
The observed dSph seem to be composed of two subsets: 
a sizeable fraction of them behave as the simulated satellites while the remainder align at
$\gtot\simeq 3\times 10^{-11}\,\mss$.

A quantitative analysis is presented in 
Fig. \ref{fig:likelihood} where we compare
the best-fit RARs for our MCGs and MSGs.
The posterior probability densities of the 
model parameters show that centrals and satellites
follow a RAR characterized by the same $\gcrit$
but with very different values for the intrinsic scatter.
In fact, for the MCGs, we find $\gcrit= (1.40 \pm 0.07)\times 10^{-10}\,\mss$ and $\sint=0.048 \pm 0.005\,\rm dex$ while, for
the satellites,
$\gcrit= (1.48 \pm 0.08)\times 10^{-10}\,\mss$ and $\sint=0.192 \pm 0.008\,\rm dex$.
The characteristic acceleration we measure is larger than, but compatible with, the observed value for MCGs which is plagued with a relatively large systematic error. We note
that the model-fitting method influences the result. For instance,
adopting the (frequentist) orthogonal-distance regression algorithm 
to fit only the characteristic acceleration (as in 
\cite{McGaugh+2016,Lelli+2017}) yields
$\gcrit = (1.19 \pm 0.02) \times 10^{-10} \, \mss$ for MCGs, 
in very good agreement with the observational results. 
Following \cite{Lelli+2017}, we define a
`high-quality' sample of satellites that contain a large number of stellar particles, have small ellipticities and are barely
affected by the tidal field of the host galaxy.
This does not significantly change the best-fit intervals for $\gcrit$ and $\sint$.

\begin{figure}
	\includegraphics[width=\columnwidth]{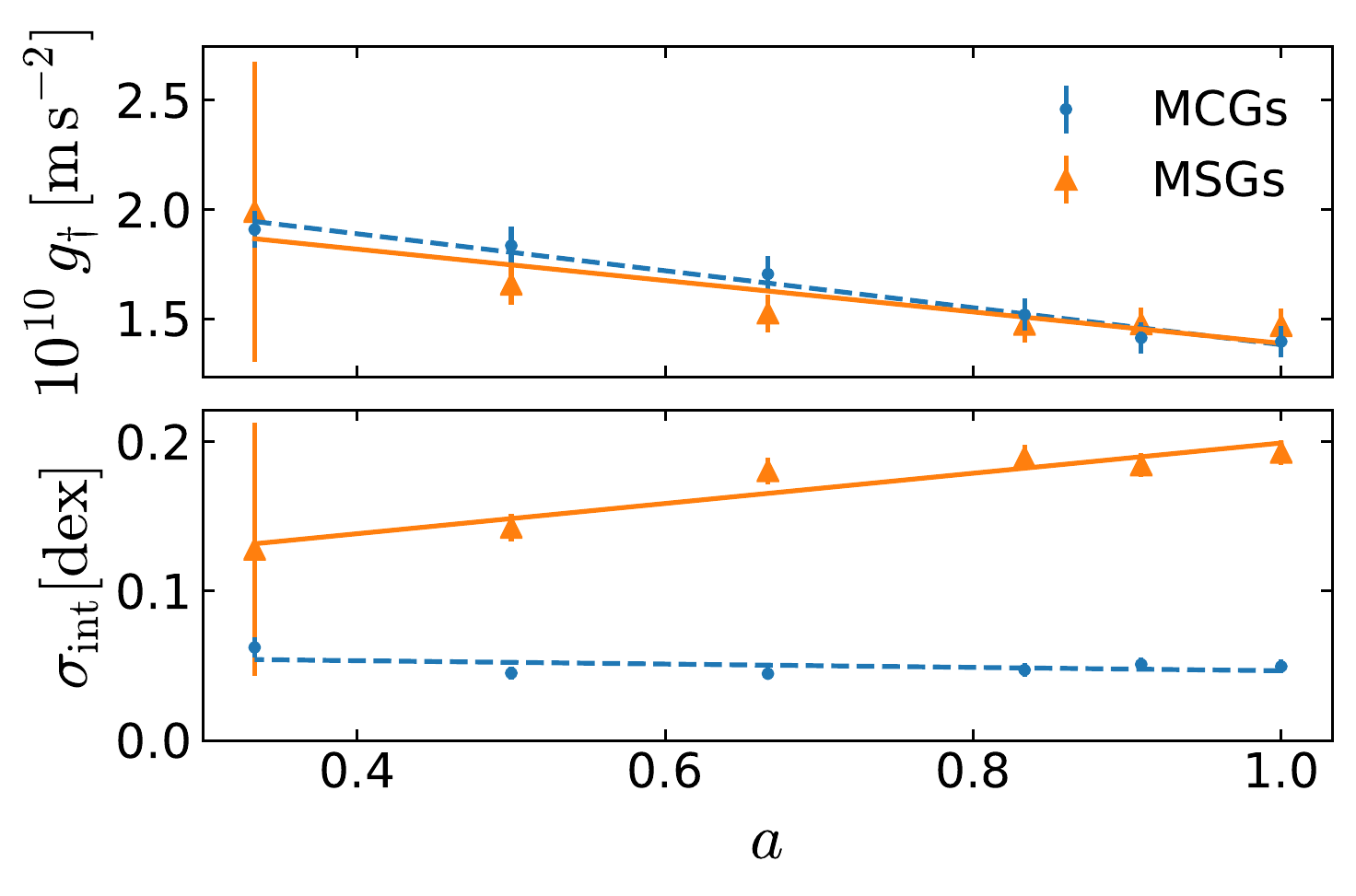}
    \caption{Evolution of $\gcrit$ (top) and $\sint$ (bottom) as a function of the expansion factor
    of the universe for MCGs (dots) and MSGs
 (triangles). The lines show the best-fit linear relations.}
    \label{fig:rar_combined}
\end{figure}

{\it Independency of the RAR on the satellite properties.{\bf ---}} 
Given the large scatter characterizing the RAR for MSGs,  
we investigate whether sub-classes of satellites with different physical properties follow distinct RARs at $z=0$. 
We first sort
the satellites based on some 
physical property. Then we separately
fit Eq.~(\ref{eq:rar}) to the subsets containing the upper and lower 20 per cent of the sorted data.
Specifically, we examine the following variables: 
(a) the tidal acceleration at $\Rhl$ due to the gravitational field
of the host galaxy, $g_{\rm tides}=2\,G\,M_{\rm host}\,\Rhl/D^3_{\rm host}$,
as defined in \cite{Lelli+2017}; 
(b) The distance of the satellite from the main galaxy;
(c) The triaxiality parameter of the stellar distribution;
(d) The minor-to-major and medium-to-major axis ratios;
(e) The cosine of the angle between the satellite velocity and the radial direction with respect to the central host;
(f) The stellar concentration defined as the ratio between the radius enclosing 80\% of the stellar mass and that enclosing 20\% of it;
(g) The accretion time of the satellite on to its host; 
(h) The mass loss experienced between
accretion time and redshift zero.
The only significant discrepancy we find is 
between the credibility intervals of 
$\sint$ for the subsamples of case (h): the scatter
is three times larger for satellites that experienced a large mass loss.

{\it Time evolution of the RAR.{\bf ---}}
Finally,
we study the RAR at $z>0$.
At all epochs, we identify a well defined 
relation
for both
MCGs and MSGs which we fit using
Eq.~(\ref{eq:rar}).
Our findings,
summarized in Fig.~\ref{fig:rar_combined}, show 
that both $\gcrit$ and $\sint$ 
evolve little with time.
To good approximation, the best-fit parameters for the RAR scale linearly with the scale factor $a$ of the universe.
In the range 
$0.33\leq a\leq 1$,
$\gcrit \simeq (-0.84 \, a + 2.23)\times 10^{-10} \, \mss$ for MCGs and
$\gcrit \simeq (-0.72 \, a + 2.11)\times 10^{-10} \, \mss$ for MSGs (the uncertainty on the parameters is $\sim10\%$).
On the other hand,
the intrinsic scatter around the RAR stays
approximately constant for MCGs,
$\sint \simeq -0.01 \, a + 0.06$ dex, 
and grows as
$\sint \simeq 0.1 \, a + 0.1$ dex for the satellites.

The evolution of the RAR for central galaxies is promoted by stellar feedback
which drives important outflows at
high redshift
\cite{Keller+Wadsley2017}.

\begin{figure}
	\includegraphics[width=\columnwidth]{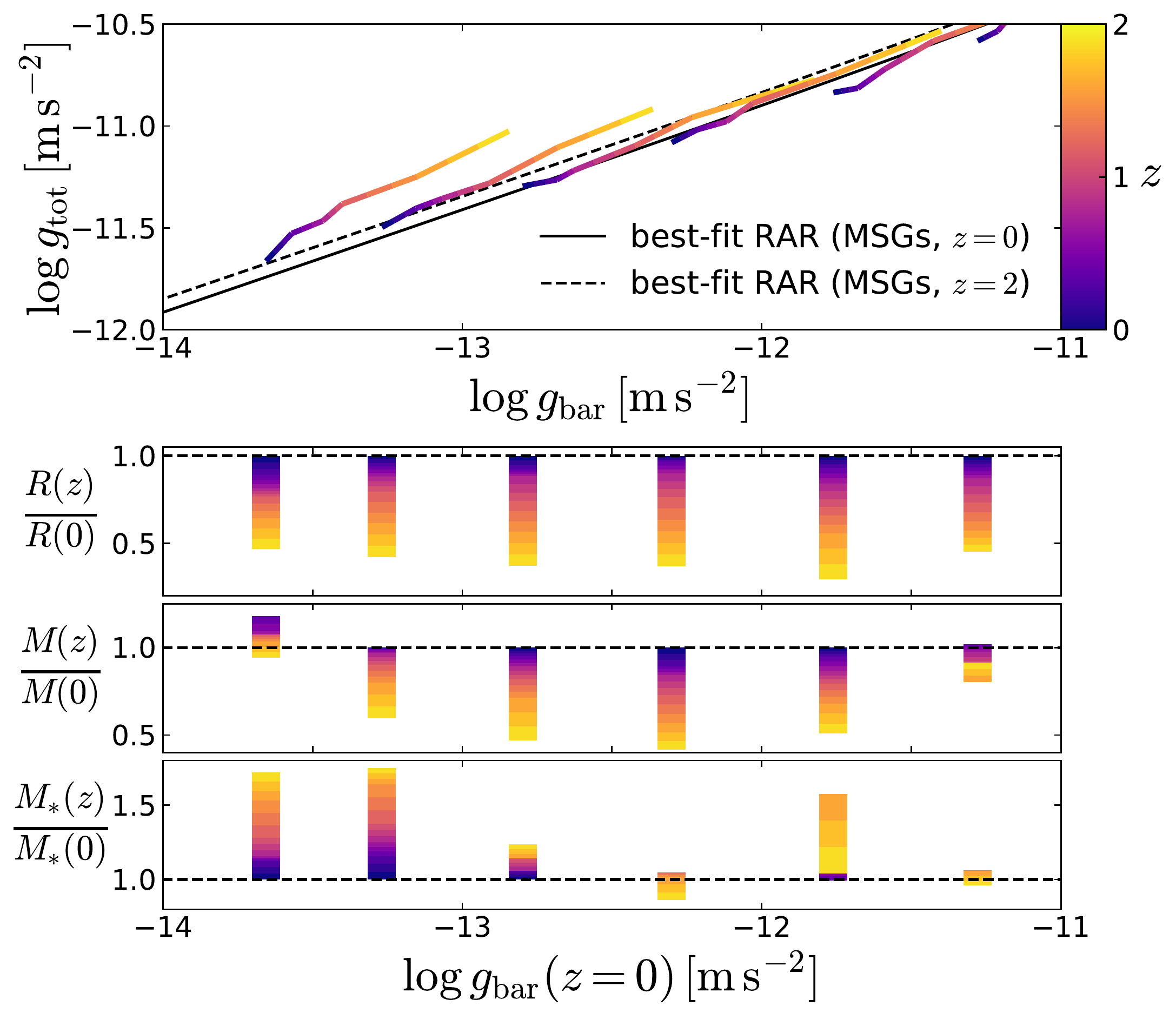}
    \caption{Top: Characteristic evolutionary tracks of MSGs obtained by partitioning
    the objects based on their value of $\gbar$
    at $z=0$ and plotting the median
    values of $\gbar$ and $\gtot$ in each bin at some earlier epoch indicated by the color scale.
    Bottom: Evolution of $\Rhl$ and of the enclosed total and stellar masses for the same bins.
    Note that each satellite is tracked from the moment it accretes on to
    its host halo to $z=0$. Therefore the number of objects in each bin decreases with increasing $z$.}
    \label{fig:track}
\end{figure}

In order to characterize the time evolution of the satellites, in the top panel of Fig.~\ref{fig:track},
we partition them based on 
their $\gbar$ at the present time and
plot the median trajectory of each
subset in the $\gbar$-$\gtot$ plane as a function of redshift (indicated by the color).
The trend is to move from the top right to the bottom left nearly parallel to the RAR. 
The other panels of Fig.~\ref{fig:track} reveal the reason for this tendency.
Essentially, while $\Rhl$ and the DM mass within it tend to grow
with time, the stellar mass of the satellites decreases.
This is the net result of tidal stripping that
makes satellites more DM dominated with time.
Since the DM and the stars in a satellite follow distinct spatial distributions at the accretion time, they react differently to tidal forces. 
The (physical) extension of the
stellar distribution increases during the evolution \citep{Penarrubia+2008,Penarrubia+2009}
while the DM density profile becomes more concentrated \citep{Libeskind+2010,RomanoDiaz+2010}.
Of course, individual objects follow 
complex trajectories in the $\gbar$-$\gtot$ plane which produce some scatter around the median trend (see also \cite{Fattahi+2017}).

{\it Conclusions.{\bf ---}}
The RAR is an empirical law describing a tight relation
between the radial acceleration generated by the visible matter in galaxies and the actual acceleration derived
from kinematic measurements. For bright central galaxies,
the correlation is such that both $\gbar$ and $\gtot$
decrease in the outer regions.
This result could hint towards a scenario in which
there is no DM and the law of gravity needs
to be modified along the lines of MOND.
Galaxy-formation models within the \LCDM scenario are able to reproduce the observed relation, although with too large a scatter.
Ref.~\cite{Lelli+2017}
provides evidence that
nearby dSph satellite galaxies depart from the RAR and show a constant $\gtot$ for $\gbar \lesssim 10^{-12} \, \mss$. 
However,
the authors caution
that unresolved
binary stars and out-of-equilibrium dynamics could bias the measurements of $\gtot$ high in these low-mass structures.
It is yet unclear what are the implications for the theory of gravity.
The missing pieces of the puzzle are (a) more precise
measurements and (b) accurate theoretical predictions
for the behaviour of satellite galaxies in \LCDM.
This work supplies the latter by making use
of a suite of zoom hydrodynamical simulations. 
Our main results are:
(i) At $z=0$, the simulated satellites scatter around a well defined sequence in the $\gbar$-$\gtot$ plane which is approximately aligned with the
observed RAR for central galaxies and does not show any transition to a constant $\gtot$ at low accelerations.
(ii) For the least massive objects, the satellite sequence is shallower than the RAR for the central galaxies. In fact, $\gtot$ scales as $\gbar^{0.4}$.
This flattening is even more prominent for dwarf
galaxies that are not satellites.
(iii) The scatter around the satellite sequence is approximately four times larger than for the central galaxies. 
(iv) Although the deviations from the main sequence do not correlate with many physical properties of the satellites, the intrinsic scatter around the RAR is three times larger for objects that were stripped off more mass.
(v) The RAR for central galaxies shows a mild evolution with redshift. 
The characteristic acceleration
decreases with time, meaning that galaxies are relatively
more baryon depleted at high redshifts with respect to
the present epoch. The scatter around the relation 
stays constant with time.
(vi) Individual satellites tend to evolve along the
$\gbar$-$\gtot$ sequence. This trend is driven by
tidal stripping combined with an internal readjustment of
the structures. Typically,  the stellar profile broadens out and $\Rhl$ increases with time
while the DM distribution gets more concentrated.
(vii) Since satellites follow the RAR of the central galaxies before accreting
on to their hosts and evolve along the main sequence
afterwards, their $\gcrit$ shows the same time evolution
as for the central galaxies.
Given the wide variety of the evolutionary paths, the scatter around the relation between the accelerations for the satellites increases with time
and with decreasing $\gbar$.
(viii) In our simulations, residuals from the RAR  for the satellites do not correlate with $g_{\rm tides}$.
Conversely, in the MOND framework, satellites in a strong external gravitational field show different internal accelerations than if they were isolated. Detecting the absence or presence of the correlations from observations would therefore provide a powerful test of the theory of gravity.

\vspace{0.2cm}
We thank Aaron Ludlow for discussions and the anonymous referees for useful comments.
This work is carried out within the SFB 956 
\virg{The Conditions and Impact of Star Formation}, sub-project C4, and the Transregio 33 `The Dark Universe' projects funded by the Deutsche Forschungsgemeinschaft (DFG). The results presented were achieved employing computing resources (Cartesius) at SURF/SARA, The Netherlands as part of the
PRACE-3IP project (FP7 RI-312763). We are thankful to the community
developing and maintaining software packages extensively used
in our work, namely: Matplotlib \citep{matplotlib},
NumPy \citep{numpy}. MSP acknowledges that support for this work was provided by NASA through Hubble Fellowship grant \#HST-HF2-51379.001-A awarded by the Space Telescope Science Institute, which is operated by the Association of Universities  for  Research  in  Astronomy,  Inc.,  for  NASA,  under  contract  NAS5-26555.

\bibliographystyle{apsrev4-1} 
\bibliography{rar_dSph} 

\label{lastpage}

\end{document}